\documentclass[jkps,preprint,fleqn,showpacs,showkeys,nofootinbib]{revtex4}
\usepackage{graphicx}
\usepackage{amssymb}
\usepackage{amsmath}
\usepackage{bm}
\begin{document}
\setcounter{page}{0}
\title{Riemannian and non-Riemannian extensions of geometrodynamics versus Einsteinian gravity}
\author{Orchidea Maria \surname{Lecian}}
\email{lecian@icra.it}
\thanks{Fax: +39-06-4454-992}
\affiliation{Universit\`a La Sapienza, Dipartimento di Fisica and ICRA, Piazzale Aldo Moro, 5-00185 Roma, Italy}
\author{Giovanni \surname{Montani}}
\affiliation{Universit\`a La Sapienza, Dipartimento di Fisica and ICRA, Piazzale Aldo Moro, 5-00185 Roma, Italy\\
ENEA -- C.R. Frascati (Department F.P.N.), Via 
Enrico Fermi, 45-00044, Frascati (Roma), Italy\\
ICRANet -- C. C. Pescara, Piazzale della 
Repubblica, 10-65100, Pescara, Italy}

\date[]{Received 6 August 2007}

\begin{abstract}
We analyze some extensions of General Relativity.\\
Within the framework of modified gravity, the Newtonian limit of a class of gravitational actions is discussed on the basis of the corresponding scalar-tensor model.\\
For a generalized asymmetric metric, autoparallel trajectories are defined under suitable conditions at first approximation order.
\end{abstract}



\maketitle

\section{INTRODUCTION}

The investigation of possible extensions of General Relativity has inspired a large amount of work. On the one hand, it is possible to preserve the metric structure of Einsteinian space time and to generalize the gravitational Lagrangian by replacing the Ricci scalar with a function of it. As a result, modified gravity can be shown to be equivalent to a scalar-tensor scheme, i.e., to a scalar field minimally coupled to gravity, via a suitable conformal transformation. On the other hand, non-Riemannian geometry can be taken into account: in the most general setting, an asymmetric metric tensor, torsion and non-metricity tensor arise. These objects lead to a modification of the gravitational Lagrangian, and their coupling with spinors can be analyzed.\\
In this paper we will address both points of view.\\
In the second section, we will analyze the implications of some relevant examples in modified gravity and the corresponding scalar-tensor model. In fact, the analysis of an exponential action for the gravitational field will outline the puzzle of vacuum energy, while the investigation of a gravitational Lagrangian consisting of positive and negative powers of the Ricci scalar will illustrate how such modified-gravity theories can mimic the effect of dark energy. On the converse, we will also investigate the Newtonian limit of some potentials, whose scalar-tensor description is explicitly solvable with respect to the first derivative.\\
In the third section, we will examine an extended Shouten classification of non-Riemannian geometries: the curvature tensor will be shown to decouple into a Riemannian part plus other contributions, and autoparallel trajectories will be evaluated at first-order approximation. The coupling between spinors and non-Riemannian objects will also be considered in the case of a symmetric metric, where torsion and non-metricity are taken into account.\\
Brief concluding remarks follow.

\section{MODIFIED GRAVITY}
Within the framework of modified gravity, when the Ricci scalar $R$ is replaced with a generic function $f(R)$, the new gravitational action, in the Jordan frame, reads
\begin{equation}\label{fdr}
S_{G}=\frac{1}{k^{2}}\int d^{4}x \sqrt{-g} f(R)
\end{equation}
whose variation with respect to $g^{\mu\nu}$ yields generalized Einstein equations. After introducing two Lagrange multipliers and eliminating one of them, (\ref{fdr}) rewrites
\begin{equation}\label{elimination}
S=\frac{1}{k^{2}}\int d^{4}x\sqrt{-g}\left[f'(A)(R-A)+f(A)\right],
\end{equation}
with $R=A$. By means of the conformal transformation
\begin{equation}\label{conf}
g_{\mu\nu}\rightarrow e^{\phi}g_{\mu\nu},
\end{equation}
and for the particular (on shell) choice $\sqrt{3/2}\phi=-\ln f'(A)$, (\ref{elimination}) is mapped into the action of  scalar field in curved space time minimally coupled to gravity \cite{intro}, whose evolution, under the on-shell condition, is given by the potential $V(\phi)$  
\begin{equation}\label{potential}
V(\phi)=\frac{A}{f'(A)}-\frac{f(A)}{f'(A)^{2}}.
\end{equation}
According to the choice of the function $f(R)$, different features of the modern universe can be explained. In fact, many open problems at different scales seem to be solved without introducing dark energy or dark matter \cite{mof}. Constraints on the free parameters of such theories \cite{mof2} can be imposed on the basis of solar-system data, binary-pulsar tests and cosmological observations \cite{neno}.\\
In particular, a crucial role is played by the value of $f(0)$: if this value is non-vanishing, a vacuum-energy cancellation mechanism has to hypothesized. The analysis of an exponential function of the Ricci scalar\cite{lm07}, i.e., $f(R)=\lambda e^{\mu R}$, illustrates that the
geometrical components contain a cosmological term too, and the deSitter solution exists in
presence of matter only for a negative ratio between the vacuum-energy density and that of the intrinsic cosmological term $\epsilon_{\Lambda}$. A negative value of the intrinsic cosmological constant predicts an accelerating deSitter dynamics, but, in this case, we would get a vacuum-energy density greater than the modulus of the intrinsic term. The vacuum solution describes the full non-perturbative regime, from which the Einsteinian limit cannot be recovered, and the introduction of the external field can be interpreted as the role dark energy plays in the vacuum-energy cancellation mechanism. The corresponding scalar-tensor model consists of s scalar field minimally coupled with gravity, evolving under a potential $V(\phi)\propto \epsilon_{\Lambda}e^{\phi}(\phi+1)$: if the sign of the intrinsic cosmological constant is reverted, this potential admits a minimum, around which field equations can be linearized. The value $\phi_0$ that minimizes the potential does not correspond to the vacuum solution of the Jordan frame, but to the solution found in presence of external matter. This achievement establishes an off-shell correspondence between the two frames, which supports the interpretation of such an unphysical external field with the action of dark energy in the vacuum-energy cancellation. \\      
The class of functions $f(R)=f_{0}R^n$ has been widely investigated\cite{cct06}, and can be considered as an eligible candidate to solve both dark-energy and dark-matter problems on cosmological and galactic scales, respectively.  In the weak-field limit, the gravitational potential reads
\begin{equation}
\Phi(r)=-\frac{Gm}{2r}\left[1+\left(\frac{r}{r_c}\right)^\beta\right],\ \ \beta=\frac{12n^2-7n-1-\sqrt{36n^4+12n^3-83n^2+50n+1}}{6n^2-4n+2}:
\end{equation}
$r_c$ is an arbitrary parameter, depending on the typical scale of the considered system, while $\beta$ is a universal parameter. In particular, the Newtonian potential is recovered for $n=1$ and $\beta=0$, while $\beta\rightarrow1$ for $n\rightarrow\infty$. On the one hand, observational data on spiral-galaxy rotation curves are fitted, within this model, by $n>1$, but, on the other hand, solar-system data based on light-bending and planetary orbits \cite{zak06} require $\beta\sim0$, which is a very stringent constraint\footnote{Anyhow, it has been demonstrated \cite{noj03} that the more general function $f(R)=R-\frac{c}{(R-\Lambda_1)^n}+b(R-\Lambda_2)^m$ can pass solar-system tests after some fine tuning.}.\\
By means of the  conformal transformation (\ref{conf}), it is found that the class of functions $f(R)=f_{0}R^n$ is mapped into a scalar field minimally coupled to gravity, with an exponential potential \cite{capozexp}, in absence of external matter. It is worth remarking that also inflationary scenarios are often realized by considering a scalar field with a particular potential, and the conformal equivalence (\ref{conf}) is the link between the two models.\\
We are now ready to go the other way round in order to investigate the expression of some potentials in the Jordan frame, by solving (\ref{potential}) for the unknown $f(R)$. It is worth noting that the condition $f''\neq 0$ has to be imposed to keep results consistent from a mathematical point of view; it automatically rules Einsteinian gravity (plus a cosmological constant) out of our investigation. (The condition $f\neq0$ excludes trivial solutions).\\ 
The preliminary analysis of the simple case of a quadratic potential \cite{quad}, $V(\phi)=m\phi^2$, restricts the range of potentials that can be studied in this perspective. In fact, in this case, (\ref{potential}) reads
\begin{equation}
f'R-f=\frac{3}{2}mf'^{2}(\lg f')^2:
\end{equation}
differentiating with respect to $R$ and introducing the auxiliary function $f'=p$, one obtains the system
\begin{align}
&R=3m(\lg p)\left(p+\lg p\right),\nonumber\\ 
&f=3mp\left(1+\frac{p}{4}\right)\left(2\lg p-1\right)+C,
\end{align}
where $C$ is an arbitrary integration constant. From this example, we can see that there is a large class of inflationary potentials \cite{mix}, which can not be solved with respect to the first derivative explicitly, and are therefore unsuitable for our investigation. Other approaches will be addressed elsewhere \cite{prep}.\\
However, the exponential case \cite{capozexp} offers a great variety of applications. For
\begin{equation}\label{espo} 
V(\phi)=\alpha e^{-\sqrt{\frac{3}{2}}\lambda\phi}, 
\end{equation}
(\ref{potential}) rewrites
\begin{equation}\label{exppot}
f'R-f=\alpha f'^{2}e^{\lambda\lg f'}\equiv \alpha f'^{q},
\end{equation}
with $q\equiv \lambda+2$: differentiating with respect to $R$, we find the general solution
\begin{equation}\label{math}
f=\frac{1}{(\alpha q)^{\frac{1}{q-1}}}\frac{q-1}{q}R^{1+\frac{1}{q-1}}+C=\frac{1}{(\alpha q)^{\frac{1}{q-1}}}\frac{q-1}{q}R^{\frac{q}{q-1}}+C,
\end{equation}
where $C$ is an arbitrary integration constant: we recover the original potential only for $C\equiv0$. This solution is consistent for $q\neq 0$ and $q\neq 1$, i.e., $\lambda\neq-2$ and $\lambda\neq-1$, respectively. This solution is also consistent with the condition $f''\neq0$, as $\frac{q}{q-1}\neq1\ \ \forall q$, and a linear gravitational Lagrangian, excluded by construction, is not a general solution for (\ref{exppot}).\\
Contrastingly, it's worth remarking that the special case $\lambda=0$, i.e., a constant potential $V(\phi)=\alpha$, is provided by both a linear gravitational Lagrangian plus a cosmological constant \cite{lambda}, and $f=R^{2}/(4\alpha)$.\\
Finally, the trivial case $V(\phi)=0$, obtained for $\alpha=0 \ \ \forall \lambda$, is provided only by a linear gravitational Lagrangian without a cosmological constant.\\ 
We can recover information about the cases $\lambda=-2$, $\lambda=-1$, $C\neq 0$ by studying the slightly different class of potentials \cite{cosh} 
\begin{equation}\label{exp2}
V(\phi)=\alpha_{1}e^{-\sqrt{\frac{3}{2}}\lambda_{1}\phi}+\alpha_{2}e^{-\sqrt{\frac{3}{2}}\lambda_{2}\phi}.
\end{equation}
Potential (\ref{exp2}) is not in general solvable with respect to the first derivative explicitly, but admits the formal solution
\begin{align}
&R=\alpha_{1}(2+\lambda_{1})p^{1+\lambda_{1}}+\alpha_{2}(2+\lambda_{2})p^{1+\lambda_{2}},\nonumber\\ &f=\alpha_{1}(1+\lambda_{1})p^{2+\lambda_{1}}+\alpha_{2}(1+\lambda_{2})p^{2+\lambda_{2}} +C,
\end{align}
where, as above, $C$ is an arbitrary integration constant, $p\equiv f'$, and $f''\neq 0$.\\
Some particular cases admit nevertheless an explicit solution. In fact, for $\lambda_{2}=-2$, we find 
\begin{equation}
f=\frac{1}{(\alpha_{1} r)^{\frac{1}{r-1}}}\frac{r-1}{r}R^{1+\frac{1}{r-1}}+\alpha_{2},
\end{equation}
while, for $\lambda_{2}=-1$, we obtain
\begin{equation}
f=\frac{1}{(\alpha_{1} r)^{\frac{1}{r-1}}}\frac{r-1}{r}(R-\alpha_{2})^{1+\frac{1}{r-1}},
\end{equation}
where, as above, $f''\neq0$ and $r\equiv 2+\lambda_{1}$: $r\neq 0$ and $r\neq 1$, i.e., $\lambda_{1}\neq-2$ and $\lambda_{1}\neq-1$, respectively.\\ 
For $\lambda_{1}=0$ of (\ref{exp2}), (\ref{potential}) is not solvable explicitly, and, differently from the previous result, a linear gravitational Lagrangian is not a solution.\\
Particular cases of (\ref{exp2}) can be analyzed. \\
For $\lambda_{1}=-\lambda_{2}=2\equiv\lambda$, we find the solution $f=\frac{1}{(4\alpha_{1})^{1/3}}\frac{3}{4}R^{4/3}+\alpha_{2}$, while, for $\lambda_{1}=-\lambda_{2}=1\equiv\lambda$, $f=\frac{1}{\sqrt{3\alpha_{1}}}\frac{2}{3}(R-\alpha_{2})^{3/2}$. For $\alpha_{1}=\alpha_{2}\equiv\alpha$, the hyperbolic-cosine potential \cite{cosh} \cite{staro}is recovered.\\
The Newtonian limit of such scenarios can be examined. In fact, for (\ref{espo}), the limit $q\rightarrow\infty$, i.e., $\lambda\rightarrow\infty$, induces a vanishing potential, and solution (\ref{math}) gives the proper limit, $f\propto R$, i.e., a linear gravitational Lagrangian without a cosmological constant, which is also achieved by imposing $\alpha\equiv0$ in (\ref{math}). From a physical point of view, for $\lambda>>1$, (\ref{math}) reduces to $f(r)\propto R^{1+\epsilon(\lambda)}$, with $\epsilon(\lambda)=1/(1+\lambda)<<1$, and (\ref{espo}) becomes negligible for small values of $\phi$. Furthermore, as discussed above, there are two classes of functions that can reproduce a constant potential, i.e., both a linear gravitational Lagrangian plus a cosmological constant and $f=R^{2}/(4\alpha)$. Anyhow, a quadratic gravitational Lagrangian can be shown to be inappropriate to fit solar-system data \cite{zak06}.

\section{NON-RIEMANNIAN GEOMETRY}
A generalized Schouten classification \cite{clmrz} can be constructed for the case of an asymmetric metric tensor $g_{\mu \nu }=s_{\mu \nu }+a_{\mu \nu }$, which splits into its symmetric and antisymmetric part, respectively. The corresponding connection $\Pi
_{\kappa \lambda }^{\theta }$ is determined by the incompatibility between metric and connection. The connection explicitly reads
\begin{equation}
\Pi _{\kappa \lambda }^{\theta }J_{\theta \nu \rho }^{\sigma
\kappa \lambda }\equiv\Pi _{\kappa \lambda }^{\theta }(\delta _{\theta }^{\sigma }\delta _{\nu
}^{\kappa }\delta _{\rho }^{\lambda }+g^{\sigma \lambda }\delta _{\rho
}^{\kappa }a_{\theta \nu }+g^{\sigma \kappa }\delta _{\nu }^{\lambda
}a_{\rho \theta })=\Gamma _{\nu \rho }^{\sigma }+\Delta _{\nu \rho }^{\sigma
}+{\bf C}_{\nu \rho }^{\sigma }-{\bf D}_{\nu \rho }^{\sigma }\,,
\label{eqndet}
\end{equation}
where $\Gamma _{\nu \rho }^{\sigma }$ is 
the usual Christoffel connection, while
\begin{subequations}\label{lkjh}
\begin{align}
&\Delta _{\nu \rho
}^{\sigma }=\frac{1}{2}s^{\sigma \mu }(a_{\mu \nu ,\rho }+a_{\rho \mu ,\nu
}-a_{\nu \rho ,\mu })
\\
&C_{\nu \rho
}^{\sigma }=\frac{1}{2}\left[ s^{\sigma \mu }(T_{\nu \mu }^{\varepsilon
}g_{\varepsilon \rho }+T_{\rho \mu }^{\varepsilon }g_{\varepsilon \nu
})+T_{\nu \rho }^{\varepsilon }g_{\varepsilon }^{.\sigma }\right] 
\\
&D_{\nu \rho }^{\sigma }=\frac{1}{2}s^{\sigma \mu
}(N_{\mu \nu \rho }+N_{\rho \mu \nu }-N_{\nu \rho \mu })
\end{align}
\end{subequations}
are the metric-asymmetricity object, the generalized contortion tensor, written as a function of the torsion field $T_{\rho \mu }^{\varepsilon }$, and the non-metricity tensor, written a a function of the covariant derivatives of the metric tensor with respect to the metric part of the connection $g_{\mu \nu \mid \rho }=N_{\mu
\nu \rho }$, respectively. $J_{\theta \nu \rho }^{\sigma\kappa \lambda }$ is the structure matrix.\\
The generalized curvature tensor $R^{\alpha }{}_{\beta \rho \sigma }$ splits into the Riemannian curvature tensor plus other contributions: 
\begin{equation}
R^{\alpha }{}_{\beta \rho \sigma }=M^{\varepsilon }{}_{\nu \rho \lambda }
\widetilde{J}_{\varepsilon \beta \sigma }^{\alpha \nu \lambda }+\Sigma
^{\alpha }{}_{\beta \rho \sigma }(\widetilde{A}^{\alpha }{}_{\beta \sigma
},\Delta ^{\alpha }{}_{\beta \sigma },\widetilde{J}_{\sigma \beta \gamma
}^{\alpha \nu \rho }),  \label{structure3}
\end{equation}
where $\Sigma ^{\alpha }{}_{\beta \rho \sigma }$ is a  tensor
constructed from generalized affine deformation tensor, metric asymmetricity
object and the inverse structure matrix and their derivatives. From the physical point of
view, it means that a field theory based on such a non-Riemannian geometry
always contains Riemannian gravity (general relativity with an appropriate
Lagrangian) and extra fields as non-Riemannian (non-gravitational) effects.\\
The expression of autoparallel trajectories 
\begin{equation}\label{autop}
\frac{du^\alpha}{d\lambda}+\Pi^\alpha_{\beta\gamma}u^\beta u^\gamma=
\frac{du^\alpha}{d\lambda}+\left(\Gamma^{\sigma}_{\nu\rho}+\Delta^{\sigma}_{\nu\rho}+C^{\sigma}_{\nu\rho}-D^{\sigma}_{\nu\rho}\right)\tilde{J}^{\alpha\nu\rho}_{\sigma\beta\gamma}u^\beta u^\gamma=0.
\end{equation}
depends crucially on  $\tilde{J}$, the inverse structure matrix. The existence of the inverse structure matrix $\tilde{J}$ is related to the existence of solutions of the system of inhomogeneous
linear algebraic equations (\ref{eqndet}) for the unknowns $\Pi
_{\beta \gamma }^{\alpha }$. In the case of small asymmetric metric $\mid
a_{\mu \nu }\mid \ll \mid s_{\mu \nu }\mid $, in linear approximation, we have 
\begin{equation}\label{invlin}
\widetilde{J}_{\sigma \beta \gamma }^{\alpha \nu \rho }={\delta }_{\sigma
}^{\alpha }{\delta }_{\beta }^{\nu }{\delta }_{\gamma }^{\rho }-s^{\alpha
\nu }{\delta }_{\beta }^{\rho }a_{\gamma \sigma }-s^{\alpha \rho }{\delta }
_{\gamma }^{\nu }a_{\sigma \beta }.
\end{equation} 
In the case of small asymmetric metric $\mid
a_{\mu \nu }\mid \ll \mid s_{\mu \nu }\mid $, and under the assumption $a\sim T\sim N$, we find, at first order,
\begin{equation}
\frac{du^\alpha}{d\lambda}+\Gamma^{\alpha}_{\nu\rho}u^\nu u^\rho+\frac{1}{2}s^{\mu\nu}\left(T^{\epsilon}_{\nu\mu}s_{\epsilon\rho}
+T^{\epsilon}_{\rho\mu}s_{\epsilon\nu}\right)-\frac{1}{2}s^{\sigma\mu}\left(N_{\mu\nu\rho}+N_{\rho\mu\nu}-N_{\nu\rho\mu}\right)\delta^{\alpha}_{\sigma}u^{\rho}u^{\sigma}=0.
\end{equation}
It's interesting to notice that torsion and the non-metricity tensor contribute at this approximation order, while the metric-asymmetricity object provides a negligible contribution.\\
The introduction of spinors in non-flat space time requires the introduction of covariant derivatives in order to restore invariance of the Dirac equation. In fact, in non-flat space time, the ordinary derivative of $\gamma^\mu$ matrices does not vanish, i.e., $\partial_{\mu} \gamma^{\nu}\neq 0$, and a covariant derivative $D_\mu$ must be found, such that $D_\mu \gamma^\nu=0$. Such covariant derivatives are obtained by introducing the connection $\Gamma_{\mu}$, such that
\begin{equation}\label{gatto}
D_{\mu}A=\nabla_{\mu}A-[\Gamma_{\mu},A]
\end{equation}
where $A$ is a generic object, $\nabla_{\mu}$ denotes the geometrical covariant derivative and $\Gamma_{\mu}$ is the spin connection. If $A$ is not a spinor, then $[\Gamma_{\mu},A]=0 \Rightarrow D_{\mu}A=\nabla_{\mu}A$, while, for spinors, (\ref{gatto}) specifies as
\begin{equation}
D_{\mu}\psi = \partial_{\mu}\psi -\Gamma_{\mu}\psi, \ \ D_{\mu}\overline{\psi} =\partial_{\mu}\overline{\psi}+\overline{\psi}\Gamma_{\mu}.
\end{equation}
In curved space time, we have
\begin{equation}
\Gamma_{\mu} = -\frac{1}{4}\gamma_{\nu}\nabla_{\mu}\gamma^{\nu}=\frac{1}{2}R_{ab\mu}\Sigma^{ab}, \ \ R_{ab\mu}\equiv R_{abc}e^{c}_{\mu},
\end{equation}
where $\Sigma^{ab}$ are the generators of the Lorentz group, $e^{a}_{\mu}$ are bein vectors, and $R_{abc}$ are the Ricci coeficients. In presence of torsion, we find
\begin{equation}
\Gamma_{\mu}=\frac{1}{2}R_{ab\mu}\Sigma^{ab}-\frac{1}{2}C_{ab\mu}\Sigma^{ab},
\end{equation}
where $C_{ab\mu}$ is a suitable bein projection of the contortion tensor. For a possible interpretation of the introduction of torsion, see, for example, \cite{tors}. In presence of non-metricity, spin connection are investigated in \cite{popla}, where it is established that the connection is antisymmetric in first two indices only for a metric-compatible affine connection, and projective invariance of the spin connection allows for the introduction of gauge fields interacting with spinors.

\section{CONCLUSIONS}
We have reviewed the two main research lines that allow for a generalization of Einsteinian gravity.\\
Within the framework of modified gravity, if the Ricci scalar is replaced by an arbitrary function of it, the new gravitational action can account for a wide range of phenomena that have no clear modelization in Einsteinian gravity. In particular, observational data and theoretical speculations can be reproduced without introducing dark energy. In fact, for an exponential gravitational action, a cancellation mechanism between the Plankian vacuum energy and the intrinsic geometrical term, responsible for the actual value of the cosmological constant, is necessary in the perturbative regime. For positive and negative powers of the Ricci scalar, the weak-field limit can be evaluated, and constraints on such functions can be found. Since any modified gravitational model in the Jordan frame is conformally equivalent in the Einstein frame to a scalar field in curved space time, minimally coupled with gravity and evolving under a given potential, particular scalar-field potentials can be analyzed. The resultant differential equation has been solved for an exponential potential, and information for the cases excluded in such an approach has been recovered for particular cases of the sum of two exponential potentials. The Newtonian limit of these models has been discussed, according to the results found for solar-system data.\\
Within the framework of non-Riemannian geometries, asymmetric non-metric connections provide a model that contains Riemannian gravity (general relativity with an appropriate Lagrangian) and extra fields as non-Riemannian (non-gravitational) effects. The inverse structure matrix has been found in the case of small asymmetric metric in linear approximation. The analysis of autoparallel trajectories in the hypothesis of small asymmetric metric and linear approximation reveals that torsion and the non-metricity tensor contribute at this approximation order, while the metric-asymmetricity object provides a negligible contribution. The introduction of spinors within this kind of schemes in ruled by the introduction of covariant derivatives that restore invariance of the Dirac equation, and room is left for the introduction of non-gravitational interactions.

\end{document}